\documentclass[aps,prl,twocolumn,superscriptaddress,showpacs]{revtex4-1}

\usepackage{amsmath}
\usepackage{amssymb}
\usepackage{graphicx}
\usepackage{epstopdf}

\newcommand{\eq}{\mathrm{(0)}}

\newcommand{\fl}{\mathrm{th}}
\newcommand{\tr}{\mathrm{T}}

\begin{document}

\title{Brownian thermophoresis of an antiferromagnetic soliton}

\author{Se Kwon Kim}
\affiliation{
	Department of Physics and Astronomy,
	University of California,
	Los Angeles, California 90095, USA
}

\author{Oleg Tchernyshyov}
\affiliation{Department of Physics and Astronomy, 
	The Johns Hopkins University,
	Baltimore, Maryland 21218, USA
}

\author{Yaroslav Tserkovnyak}
\affiliation{
	Department of Physics and Astronomy,
	University of California,
	Los Angeles, California 90095, USA
}

\date{\today}

\begin{abstract}
We study dynamics of an antiferromagnetic soliton under a temperature gradient. To this end, we start by phenomenologically constructing the stochastic Landau-Lifshitz-Gilbert equation for an antiferromagnet with the aid of the fluctuation-dissipation theorem. We then derive the Langevin equation for the soliton's center of mass by the collective coordinate approach. An antiferromagentic soliton behaves as a classical massive particle immersed in a viscous medium. By considering a thermodynamic ensemble of solitons, we obtain the Fokker-Planck equation, from which we extract the average drift velocity of a soliton. The diffusion coefficient is inversely proportional to a small damping constant $\alpha$, which can yield a drift velocity of tens of m/s under a temperature gradient of $1$ K/mm for a domain wall in an easy-axis antiferromagnetic wire with $\alpha \sim 10^{-4}$.
\end{abstract}

\pacs{75.78.-n, 66.30.Lw, 75.10.Hk}

\maketitle

\emph{Introduction.}|Ordered magnetic materials exhibit solitons and defects that are stable for topological reasons \cite{KosevichPR1990}. Well-known examples are a domain wall (DW) in an easy-axis magnet or a vortex in a thin film. Their dynamics have been extensively studied because of fundamental interest as well as practical considerations such as the racetrack memory \cite{ParkinScience2008}. A ferromagnetic (FM) soliton can be driven by various means, e.g., an external magnetic field \cite{SchryerJAP1974} or a spin-polarized electric current \cite{BergerPRB1996, *SlonczewskiJMMM1996}. Recently, the motion of an FM soliton under a temperature gradient has attracted a lot of attention owing to its applicability in an FM insulator \cite{HinzkePRL2011, *SchlickeiserPRL2014, BauerNM2012, YanPRL2011, *KovalevEPL2012, KongPRL2013, *KovalevPRB2014}. A temperature gradient of $20$ K/mm has been demonstrated to drive a DW at a velocity of 200 $\mu$m/s in an yttrium iron garnet film \cite{JiangPRL2013, *ChicoPRB2014}.

An antiferromagnet (AFM) is of a great current interest in the field of spintronics \cite{ZuticRMP2004, SinovaNM2012, MacDonaldPTRSA2011, *GomonayLTP2014} due to a few advantages over an FM. First, the characteristic frequency of an AFM is several orders higher than that of a typical FM, e.g., a timescale of optical magnetization switching is an order of ps for AFM NiO \cite{FiebigJPD2008} and ns for FM CrO$_2$ \cite{ZhangPRL2002}, which can be exploited to develop faster spintronic devices. Second, absence of net magnetization renders the interaction between AFM particles weak, and, thus, leads us to prospect for high-density AFM-based devices. Dynamics of an AFM soliton can be induced by an electric current or a spin wave \cite{HalsPRL2011, TvetenPRL2013, TvetenPRL2014, *KimPRB2014}. 

A particle immersed in a viscous medium exhibits a Brownian motion due to a random force that is required to exist to comply with the fluctuation-dissipation theorem (FDT) \cite{LL5, OgataJPSJ1986}. An externally applied temperature gradient can also be a driving force, engendering a phenomenon known as thermophoresis \cite{PiazzaJPCM2008}. Dynamics of an FM and an AFM includes spin damping, and, thus, involves thermal fluctuations at a finite temperature \cite{BrownPR1963, *KuboPTPS1970, *Garcia-PalaciosPRB1998, *ForosPRB2009, *HoffmanPRB2013, *AtxitiaPRB2012}. The corresponding thermal stochastic field influences dynamics of a magnetic soliton \cite{KongPRL2013, IvanovJPCM1993, SchuttePRB2014}, e.g., by assisting a current-induced motion of an FM DW \cite{DuinePRL2007}.

\begin{figure}
\includegraphics[width=0.9\columnwidth]{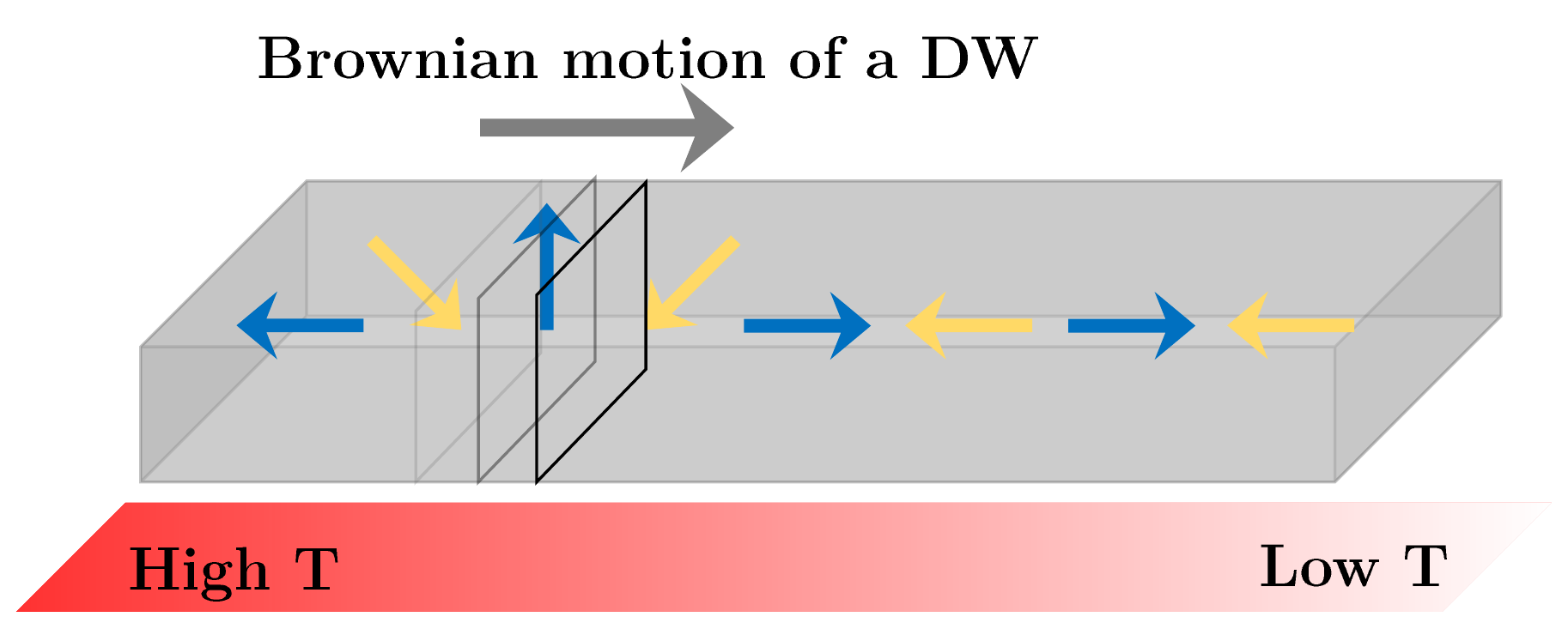}
\caption{(Color online) A thermal stochastic force caused by a temperature gradient pushes an antiferromagnetic domain wall to a colder region. The diffusion coefficient of the domain wall is inversely proportional to a small damping constant, which may give rise to a sizable drift velocity.}
\label{fig:fig1}
\end{figure}

In this Rapid Communication, we study the Brownian motion of a soliton in an AFM under a temperature gradient. We derive the stochastic Landau-Lifshitz-Gilbert (LLG) equation for an AFM with the aid of the FDT, which relates the fluctuation of the staggered and net magnetization to spin damping. We then derive the Langevin equation for the soliton's center of mass by employing the collective coordinate approach \cite{TretiakovPRL2008, TvetenPRL2013}. We develop the Hamiltonian mechanics for collective coordinates and conjugate momenta of a soliton, which sheds light on stochastic dynamics of an AFM soliton; it can be considered as a classical massive particle moving in a viscous medium. By considering a thermodynamic ensemble of solitons, we obtain the Fokker-Planck equation, from which we extract the average drift velocity. As a case study, we compute the drift velocity of a DW in a quasi one-dimensional easy-axis AFM. 

Thermophoresis of a Brownian particle is a multifaceted phenomenon, which involves several competing mechanisms. As a result, a motion of a particle depends on properties of its environment such as a medium or a temperature $T$ \cite{HottovyEPL2012}. For example, particles in protein (e.g., lysozyme) solutions move to a colder region for $T > 294\,$K and otherwise to a hotter region \cite{IacopiniEPL2003, PiazzaJPCM2008}. Thermophoresis of an AFM soliton would be at least as complex as that of a Brownian particle. We focus on one aspect of it in this Rapid Communication; the effect of thermal stochastic force on dynamics of the soliton. We discuss two other possible mechanisms, the effects of a thermal magnon current and an entropic force \cite{HinzkePRL2011}, later in the Rapid Communication.

\emph{Main results}.|Before pursuing details of derivations, we first outline our three main results. Let us consider a bipartite AFM with two sublattices that can be transformed into each other by a symmetry transformation of the crystal. Its low-energy dynamics can be developed in terms of two fields: the unit staggered spin field $\mathbf{n} \equiv (\mathbf{m}_1 - \mathbf{m}_2)/2$ and the small net spin field $\mathbf{m} \equiv (\mathbf{m}_1 + \mathbf{m}_2)/2$ perpendicular to $\mathbf{n}$. Here, $\mathbf{m}_1$ and $\mathbf{m}_2$ are unit vectors along the directions of spin angular momentum in the sublattices. 

Starting from the standard Lagrangian description of the antiferromagnetic dynamics \cite{AndreevSPU1980}, we will show below that the appropriate theory of dissipative dynamics of antiferromagnets at a finite temperature is captured by the stochastic LLG equation
\begin{subequations}
\label{eq:llg-2}
\begin{align}
s ( \dot{\mathbf{n}} + \beta \mathbf{n} \times \dot{\mathbf{m}} ) = & \mathbf{n} \times (\mathbf{h} + \mathbf{h}^\fl), \\
s ( \dot{\mathbf{m}} + \beta \mathbf{m} \times \dot{\mathbf{m}} + \alpha \mathbf{n} \times \dot{\mathbf{n}} ) 
	=	& \mathbf{n} \times (\mathbf{g} + \mathbf{g}^\fl) \nonumber \\
		&+ \mathbf{m} \times (\mathbf{h} + \mathbf{h}^\fl),
\end{align}
\end{subequations}
in conjunction with the correlators of the thermal stochastic fields $\mathbf{g}^\fl$ and $\mathbf{h}^\fl$,
\begin{subequations}
\label{eq:corr}
\begin{align}
\langle g^\fl_i (\mathbf{r}, t) g^\fl_j (\mathbf{r'}, t') \rangle &= 2 k_B T \alpha s \delta_{ij} \delta(\mathbf{r} - \mathbf{r'}) \delta(t - t'), \label{eq:g-corr} \\
\langle h^\fl_i (\mathbf{r}, t) h^\fl_j (\mathbf{r'}, t') \rangle &= 2 k_B T \beta s \delta_{ij} \delta(\mathbf{r} - \mathbf{r'}) \delta(t - t'),
\end{align}
\end{subequations}
which are independent of each other \footnote{Employing the quantum FDT would change the correlator of the stochastic fields to $\protect \langle h^\fl_i (\mathbf{r},\omega) h^\fl_j (\mathbf{r}',\omega') \protect \rangle = [ 2\pi\delta_{ij}\alpha s\hbar \omega / \tanh(\hbar \omega / 2 k_B T)] \delta(\mathbf{r} - \mathbf{r}') \delta(\omega - \omega')$ in the frequency space \cite{LL5}. We focus on slow dynamics of an AFM soliton in the manuscript, $\hbar \omega \ll k_B T$, which allows us to replace $\hbar \omega / \tanh(\hbar \omega / 2 k_B T)$ with $2 k_B T$, yielding Eq.~(\ref{eq:corr}).}. This is our first main result. Here, $\alpha$ and $\beta$ are the damping constants associated with $\dot{\mathbf{n}}$ and $\dot{\mathbf{m}}$, $\mathbf{g} \equiv - \delta U / \delta \mathbf{n}$ and $\mathbf{h} \equiv - \delta U / \delta \mathbf{m}$ are the effective fields conjugate to $\mathbf{n}$ and $\mathbf{m}$, $U[\mathbf{n}, \mathbf{m}] \equiv U[\mathbf{n}] + \int dV |\mathbf{m}|^2 / 2 \chi$ is the potential energy ($\chi$ represents the magnetic susceptibility), and $s \equiv \hbar S / \mathcal{V}$ is the spin angular momentum density ($\mathcal{V}$ is the volume per spin) per each sublattice. The potential energy $U[\mathbf{n}(\mathbf{r}, t)]$ is a general functional of $\mathbf{n}$, which includes the exchange energy $\int dV A_{ij} \partial_i \mathbf{n} \cdot \partial_j \mathbf{n}$ at a minimum \cite{AndreevSPU1980}.

Slow dynamics of stable magnetic solitons can often be expressed in terms of a few collective coordinates parametrizing slow modes of the system. The center of mass $\mathbf{R}$ represents the proper slow modes of a rigid soliton when the translational symmetry is weakly broken. Translation of the stochastic LLG equation~(\ref{eq:llg-2}) into the language of the collective coordinates results in our second main result, a Langevin equation for the soliton's center of mass $\mathbf{R}$:
\begin{equation}
M \ddot{\mathbf{R}} + \Gamma \dot{\mathbf{R}} = - \partial U / \partial \mathbf{R} + \mathbf{F}^\fl,
\label{eq:langevin}
\end{equation}
which adds the stochastic force $\mathbf{F}^\fl$ to Eq.~(5) of \textcite{TvetenPRL2014}. The mass and dissipation tensors are symmetric and proportional to each other: 
$
M_{ij} \equiv \rho \int dV (\partial_i \mathbf n \cdot \partial_j \mathbf n)
$
and
$
\Gamma_{ij} \equiv M_{ij} / \tau,
$
where $\tau \equiv \rho / \alpha s$ is the relaxation time, $\rho \equiv \chi s^2$ is the inertia of the staggered spin field $\mathbf{n}$. The correlator of the stochastic field $\mathbf{F}^\fl$ obeys the Einstein relation
\begin{equation}
\langle F^\fl_i (t) F^\fl_j (t') \rangle = 2 k_B T \Gamma_{ij} \delta(t - t').
\end{equation}

A temperature gradient causes a Brownian motion of an AFM soliton toward a colder region. In the absence of a deterministic force, the average drift velocity is proportional to a temperature gradient $V \propto k_B \nabla T$ in the linear response regime. The form of the proportionality constant can be obtained by a dimensional analysis. Let us suppose that the mass and dissipation tensors are isotropic. The Langevin equation~(\ref{eq:langevin}) is, then, characterized by three scalar quantities: the mass $M$, the viscous coefficient $\Gamma$, and the temperature $T$, which define the unique set of natural scales of time $\tau \equiv M/\Gamma$, length $l \equiv \sqrt{k_B T M} / \Gamma$, and energy $\epsilon \equiv k_B T$. Using these scales to match the dimension of a velocity yields
$
V = - c \mu (k_B \nabla T),
$
where $\mu \equiv \Gamma^{-1}$ is the mobility of an AFM soliton and $c$ is a numerical constant. The explicit solution of the Fokker-Planck equation, indeed, shows $c = 1$. This simple case illustrates our last main result; a drift velocity of an AFM soliton under a temperature gradient in the presence of a deterministic force $F$ is given by
\begin{equation}
\mathbf{V} = \mu \mathbf{F} -  \mu (k_B \boldsymbol{\nabla} T).
\end{equation}

For a DW in an easy-axis one-dimensional AFM, the mobility is $\mu = \lambda / 2 \alpha s \sigma$, where $\lambda$ is the width of the wall and $\sigma$ is the cross-sectional area of the AFM. For a numerical estimate, let us take an angular momentum density $s = 2 \hbar \, \text{nm}^{-1}$, a width $\lambda = 100 \text{ nm}$, and a damping constant $\alpha = 10^{-4}$ following the previous studies \cite{TvetenPRL2014, TakeiPRB2014}. For these parameters, the AFM DW moves at a velocity $V = 32 \text{ m/s}$ for the temperature gradient of $\nabla T = 1 \text{ K/mm}$.

\emph{Stochastic LLG equation.}|Long-wave dynamics of an AFM on a bipartite lattice at zero temperature can described by the Lagrangian \cite{AndreevSPU1980}
\begin{equation}
L = s \int dV \mathbf{m} \cdot (\mathbf{n} \times \dot{\mathbf{n}}) - U [\mathbf{n}, \mathbf{m}].
\label{eq:L}
\end{equation}
We use the potential energy $U[\mathbf{n}, \mathbf{m}] \equiv \int dV |\mathbf{m}|^2 / 2 \chi + U[\mathbf{n}]$ throughout the Rapid Communication, which respects the sublattice exchange symmetry ($\mathbf{n} \rightarrow - \mathbf{n}, \, \mathbf{m} \rightarrow \mathbf{m}$). Minimization of the action subject to nonlinear constraints $|\mathbf{n}| = 1$ and $\mathbf{n} \cdot \mathbf{m} = 0$ yields the equations of motion for the fields $\mathbf{n}$ and $\mathbf{m}$. Damping terms that break the time reversal symmetry can be added to the equations of motion to the lowest order, which are first order in time derivative and zeroth order in spatial derivative. The resultant phenomenological LLG equations are given by
\begin{subequations}
\label{eq:llg}
\begin{align}
s ( \dot{\mathbf{n}} + \beta \mathbf{n} \times \dot{\mathbf{m}} ) &= \mathbf{n} \times \mathbf{h}, \label{eq:llg-dot-n} \\
s ( \dot{\mathbf{m}} + \beta \mathbf{m} \times \dot{\mathbf{m}} + \alpha \mathbf{n} \times \dot{\mathbf{n}} ) &= \mathbf{n} \times \mathbf{g} + \mathbf{m} \times \mathbf{h}
\end{align}
\end{subequations}
\cite{TakeiPRB2014, TvetenPRL2013, IvanovPRL1994, *PapanicolaouPRB1995, *GomonayPRB2010, *SwavingPRB2011}. The damping terms can be derived from the Rayleigh dissipation function
\begin{equation}
\label{eq:R}
R = \int dV (\alpha s |\dot{\mathbf{n}}|^2 + \beta s |\dot{\mathbf{m}}|^2) / 2,
\end{equation}
which is related to the energy dissipation rate by $- \dot{U} = 2 R$. The microscopic origin of damping terms does not concern us here but it could be, e.g., caused by thermal phonons that deform the exchange and anisotropy interaction.

At a finite temperature, thermal agitation causes fluctuations of the spin fields $\mathbf{n}$ and $\mathbf{m}$. These thermal fluctuations can be considered to be caused by the stochastic fields $\mathbf{g}^\fl$ and $\mathbf{h}^\fl$ with zero mean, which  are conjugate to $\mathbf{n}$ and $\mathbf{m}$, respectively; their noise correlators are then related to the damping coefficients by the FDT. The standard procedure to construct the noise sources yields the stochastic LLG equation~(\ref{eq:llg-2}). The correlators of the stochastic fields are obtained in the following way \cite{LL5, TserkovnyakPRB2009}. Casting the linearized LLG equation~(\ref{eq:llg}) into the form $\{ \mathbf{h}, \mathbf{g} \} = \hat{\gamma} \otimes \{ \dot{\mathbf{n}}, \dot{\mathbf{m}} \}$ provides the kinetic coefficients $\hat{\gamma}$. Symmetrizing the kinetic coefficients $\hat{\gamma}$ produces the correlators~(\ref{eq:corr}) of the stochastic fields consistent with the FDT.

\emph{Langevin equation.}|For slow dynamics of an AFM, the energy is mostly dissipated through the temporal variation of the staggered spin field $\mathbf{n}$ due to $|\dot{\mathbf{m}}|^2 \simeq (\alpha \tau)^2 |\ddot{\mathbf{n}}|^2 \ll |\dot{\mathbf{n}}|^2$ (from Eq.~(\ref{eq:llg})), which allows us to set $\beta = 0$ to study long-term dynamics of the magnetic soliton \cite{TvetenPRL2014}. At this point, we switch to the Hamiltonian formalism of an AFM \cite{MikeskaJPC1980, *HaldanePRL1983}, which sheds light on the stochastic dynamics of a soliton. The canonical momentum field $\boldsymbol{\pi}$ conjugate to the staggered spin field $\mathbf{n}$ is
\begin{equation}
\boldsymbol{\pi} \equiv \delta L / \delta \dot{\mathbf{n}} = s \mathbf{m} \times \mathbf{n}.
\end{equation}
The stochastic LLG equations (\ref{eq:llg-2}) can be interpreted as Hamilton's equations,
\begin{equation}
\dot{\mathbf{n}} = \delta H / \delta \boldsymbol{\pi} = \boldsymbol{\pi} / \rho, \quad \dot{\boldsymbol{\pi}} = - \delta H / \delta \mathbf{n} - \delta R / \delta \dot{\mathbf{n}} + \mathbf{g}^\fl,
\label{eq:Hamilton-n-pi}
\end{equation}
with the Hamiltonian
\begin{equation}
H \equiv \int dV \boldsymbol{\pi} \cdot \dot{\mathbf{n}} - L = \int dV \frac{|\boldsymbol{\pi}|^2}{2 \rho} + U[\mathbf{n}].
\label{eq:H-pi-n}
\end{equation}

Long-time dynamics of magnetic texture can often be captured by focusing on a small subset of slow modes, which are parametrized by the collective coordinates $\mathbf{q} = \{q_1, q_2, \cdots \}$. A classical example is a DW in a one-dimensional easy-axis magnet described by the position of the wall $X$ and the azimuthal angle $\Phi$ \cite{SchryerJAP1974, MikeskaJPC1980}. Another example is a skyrmion in an easy-axis AFM film, which is described by the position $\mathbf{R} = (X, Y)$ \cite{RaicevicPRL2011, *ZhangarXiv2015, BarkerarXiv2015}. Translation from the field language into that of collective coordinates can be done as follows. If the staggered spin field $\mathbf n$ is encoded by coordinates $\mathbf q$ as $\mathbf{n}(\mathbf{r}, t) = \mathbf{n}[\mathbf{r}; \mathbf{q}(t)]$, time dependence of $\mathbf n$ reflects evolution of the coordinates: $\dot{\mathbf n} = \dot{q}_i \, \partial \mathbf n / \partial q_i$. With the canonical momenta $\mathbf p$ defined by 
\begin{equation}
p_i \equiv \frac{\partial L}{\partial \dot{q}_i} = \int dV \frac{\partial \mathbf n}{\partial q_i} \cdot \boldsymbol{\pi},
\end{equation}
Hamilton's equations (\ref{eq:Hamilton-n-pi}) translate into
\begin{equation}
M \dot{\mathbf q} = \mathbf p, 
\quad
\dot{\mathbf p} + \Gamma \dot{\mathbf{q}} = \mathbf{F} + \mathbf{F}^\fl,
\label{eq:Hamilton-q-p} 
\end{equation}
where $\mathbf{F} \equiv - \partial U/ \partial \mathbf q$ is the deterministic force and $F^\fl_i \equiv \int dV \partial_{q_i} \mathbf{n}  \cdot \mathbf{g}^\fl$ is the stochastic force. Hamilton's equations~(\ref{eq:Hamilton-q-p}) can be derived from the Hamiltonian in the collective coordinates and conjugate momenta, 
\begin{equation}
H \equiv \mathbf{p}^\tr M^{-1} \mathbf{p} / 2 + U(\mathbf{q}),
\label{eq:H}
\end{equation}
with the Poisson brackets $\{ q_i, p_j \} = \delta_{ij}, \, \{ q_i, q_j \} = \{ p_i, p_j \} = 0$. An AFM soliton, thus, behaves as a classical particle moving in a viscous medium. 

We focus on a translational motion of a rigid AFM soliton by choosing its center of mass as the collective coordinates $\mathbf{q} = \mathbf{R}$; $\mathbf{n}(\mathbf{r}, t) = \mathbf{n}(\mathbf{r} - \mathbf{R}(t))$. Eliminating momenta from Hamilton's equations (\ref{eq:Hamilton-q-p}) yields the Langevin equation for the soliton's center of mass:
\begin{equation}
\tau \ddot{\mathbf{R}} + \dot{\mathbf{R}} =  \mu \mathbf{F} + \boldsymbol{\eta},
\label{eq:langevin-2}
\end{equation}
where $\boldsymbol{\eta} \equiv \mu \mathbf{F}^\fl$ is the stochastic velocity. Here the mobility tensor of the soliton $\mu \equiv \Gamma^{-1}$ relates a deterministic force to a drift velocity $\langle \dot{\mathbf{R}} \rangle = \mu \mathbf{F}$ at a constant temperature \footnote{The relation between the mobility and dissipation tensor can also be understood by equating the (twice) Rayleigh function, $2 R = \mathbf{V}^\tr \Gamma \mathbf{V}$, to the dissipation governed by the mobility, $- \dot{E} = \mathbf{V} \cdot \mathbf{F} = \mathbf{V}^\tr \mu^{-1} \mathbf{V}$.}. The mobility is inversely proportional to a damping constant, which can be a small number for an AFM, e.g., $\alpha \sim 10^{-4}$ for NiO \cite{KampfrathNP2011}. The correlator~(\ref{eq:corr}) of thermal stochastic fields is translated into the correlator of the stochastic velocity,
\begin{equation}
\langle\eta_i (t) \, \eta_j (t') \rangle = 2 k_B T \mu_{ij} \delta(t - t') \equiv 2 D_{ij} \delta (t - t').
\label{eq:eta-corr}
\end{equation}

From Eq.~(\ref{eq:eta-corr}), we see that diffusion coefficient and the mobility of the soliton respect the Einstein-Smoluchowski relation: $D = \mu k_B T$, which is expected on general grounds. It can also be explicitly verified as follows. A system of an ensemble of magnetic solitons at thermal equilibrium is described by the partition function $Z \equiv \int \Pi_{i} [ dp_i dx_i / 2 \pi \hbar ] \exp( - H / k_B T)$, which provides the autocorrelation of the velocity, $\langle \dot{x}_i \dot{x}_j \rangle / 2 = M^{-1}_{ij} k_B T / 2$ (the equipartition theorem). In the absence of an external force, multiplying $\tau \ddot{x}_i + \dot{x}_i = \eta_i$~(\ref{eq:langevin-2}) by $x_j$ and symmetrizing it with respect to indices $i$ and $j$ give the equation, $\tau d^2 \langle x_i x_j \rangle / dt^2 + d \langle x_i x_j \rangle / d t = 2 \tau \langle \dot{x}_i \dot{x}_j \rangle$, where the first term can be neglected for long-term dynamics $t \gg \tau$. This equation in conjunction with the autocorrelation of the velocity allows us to obtain the diffusion coefficient $D_{ij}$ in Eq.~(\ref{eq:eta-corr}), $\langle x_i x_j \rangle = 2 k_B T \tau M^{-1}_{ij} t = 2 D_{ij} t$, without prior knowledge about the correlator~(\ref{eq:corr}) of the stochastic fields.

\emph{Average dynamics.}|An AFM soliton exhibits Brownian motion at a finite temperature. The following Fokker-Planck equation for an ensemble of solitons in an inhomogeneous medium describes the evolution of the density $\rho (\mathbf{R}, t)$ at time $t \gg \tau$:
\begin{equation}
\frac{\partial \rho}{\partial t} + \boldsymbol{\nabla} \cdot \mathbf{j} = 0, \text{ with } \mathbf{j} \equiv \mu \mathbf{F} \rho - D \boldsymbol{\nabla} \rho -  D_\text{T} (k_B \boldsymbol{\nabla} T),
\label{eq:fp}
\end{equation}
where $D_T \equiv \mu \rho$ is the thermophoretic mobility (also known as the thermal diffusion coefficient) \cite{vanKampenIBM1988, *KuroiwaJPA2014, PiazzaJPCM2008}. A steady-state current density $\mathbf{j} = \mu \mathbf{F} \rho_0 - D_\text{T} (k_B \boldsymbol{\nabla} T)$ with a constant soliton density $\rho(\mathbf{r}, t) = \rho_0$ solves the Fokker-Planck equation~(\ref{eq:fp}), from which the average drift velocity of a soliton can be extracted \cite{BraibantiPRL2008}:
\begin{equation}
\label{eq:V}
\mathbf{V} = \mu \mathbf{F} -  \mu (k_B \boldsymbol{\nabla} T).
\end{equation}

Let us take an example of a DW in a quasi one-dimensional easy-axis AFM with the energy $U[\mathbf{n}] = \int dV (A |\partial_x \mathbf{n}|^2 - K n_z^2) / 2$. A DW in the equilibrium is $\mathbf{n}^\eq = (\sin \theta \cos \Phi, \sin \theta \sin \Phi, \cos \theta)$ with $\cos \theta = \tanh[(x - X)/\lambda]$, where $\lambda \equiv \sqrt{A/K}$ is the width of the wall. The position $X$ and the azimuthal angle $\Phi$ parametrize zero-energy modes of the DW, which are engendered by the translational and spin-rotational symmetry of the system. Their dynamics are decoupled, $\Gamma_{X \Phi} = 0$, which allows us to study the dynamics of $X$ separately from $\Phi$. The mobility of the DW is $\mu = \lambda / 2 \alpha s \sigma$, where $\sigma$ is the cross-sectional area of the AFM. The average drift velocity~(\ref{eq:V}) is given by
\begin{equation}
\label{eq:V-DW}
V = - \frac{1}{2 \alpha} \frac{k_B \lambda \nabla T}{s \sigma}.
\end{equation}

\emph{Discussion}|The deterministic force $\mathbf{F}$ on an AFM soliton can be extended to include the effect of an electric current, an external field, and a spin wave \cite{HalsPRL2011, TvetenPRL2013, TvetenPRL2014}. It depends on details of interaction between the soliton and the external degrees of freedom, whose thorough understanding would be necessary for a quantitative theory for the deterministic drift velocity $\mu \mathbf{F}$. The Brownian drift velocity $\mathbf{V}$~(\ref{eq:V}) is, however, determined by local property of the soliton. We have focused on the thermal stochastic force as a trigger of thermophoresis of an AFM soliton in this Rapid Communication. There are two other possible ingredients of thermophoresis of a magnetic soliton. One is a thermal magnon current, scattering with which could exert a force on a soliton \footnote{The effect is, however, negligible at a temperature lower than the magnon gap ($\sim 40$ K for NiO \cite{KampfrathNP2011}). Also for our example---a DW in a 1D easy-axis AFM---the conservative force and torque exerted by magnons vanish \cite{TvetenPRL2014}}. The other is an entropic force, which originates from thermal softening of the order-parameter stiffness \cite{SchlickeiserPRL2014}. Effects of these two mechanisms have not been studied for an AFM soliton; full understanding of its thermophoresis is an open problem.

In order to compare different mechanisms of thermally-driven magnetic soliton motion, let us address a closely related problem of thermophoresis of a DW in a quasi one-dimensional FM wire with an easy-$xz$-plane easy-$z$-axis \cite{SchryerJAP1974}, which has attracted a considerable scrutiny recently. To that end, we have adapted the approach developed in this Rapid Communication to the FM case, which leads to the conclusion that a DW drifts to a colder region by a Brownian stochastic force at the velocity given by the same expression for an AFM DW, $V^\text{B} = - k_B \lambda \nabla T / 2 \alpha s \sigma$ \footnote{Unlike 1D domain walls, Brownian motions are drastically distinct between 2D FM and AFM solitons due to the gyrotropic force, which significantly slows down ferromagnetic diffusion \cite{SchuttePRB2014, BarkerarXiv2015}.}. A thermal magnon current pushes a DW to a hotter region at the velocity $V^\text{M} = k_B \nabla T / 6 \pi^2  s \lambda_m$, where $\lambda_m \equiv \sqrt{\hbar A / s T}$ is the thermal-magnon wavelength \cite{KovalevEPL2012}. According to \textcite{SchlickeiserPRL2014}, an entropic force drives a DW to a hotter region at the velocity $V^\text{E} = k_B \nabla T / 4 s a$, where $a$ is the lattice constant. The Brownian stochastic force, therefore, dominates the other forces for a thin wire, $\sigma \ll \lambda a / \alpha$ (supposing rigid motion) \footnote{A DW in a wire with a large crosssection $\sigma \gg a^2$ forms a 2D membrane. Its fluctuations foment additional soft modes of the dynamics, which needs to be taken into account to understand the dynamics of such a DW \cite{TretiakovPRL2008}.}. 

Within the framework of the LLG equations that are first order in time derivative, the thermal noise is white as long as slow dynamics of a soliton is concerned, i.e., the highest characteristic frequency of the natural modes parametrized by the collective coordinates is much smaller than the temperature scale, $\hbar \omega \ll k_B T$. The thermal noise could be colored in general \cite{HottovyEPL2012}, e.g., for fast excitations of magnetic systems, which may be examined in the future. In addition, local energy dissipation (\ref{eq:R}) allowed us to invoke the standard FDT at the equilibrium to derive the stochastic fields. It would be worth pursuing to understand dissipative dynamics of general magnetic systems, e.g., with nonlocal energy dissipation with the aid of generalized FDTs at the out-of-equilibrium \cite{BaiesiPRL2009, *SeifertEPL2010}.

We have studied dynamics of an AFM soliton in the Hamiltonian formalism. Hamiltonian's equations~(\ref{eq:Hamilton-q-p}) for the collective coordinates and conjugate momenta can be derived from the Hamiltonian~(\ref{eq:H}) with the conventional Poisson bracket structure. By replacing Poisson brackets with commutators, the coordinates and conjugate momenta can be promoted to quantum operators. This may provide a one route to study the effect of quantum fluctuations on dynamics of an AFM soliton \cite{LinPRB2013}.

After the completion of this work, we became aware of two recent reports. One is on thermophoresis of an AFM skyrmion \cite{BarkerarXiv2015}, whose numerical simulations support our result on diffusion coefficient. The other is on thermophoresis of an FM DW by a thermodynamic magnon recoil \cite{YanarXiv2015}.

\begin{acknowledgments}
We are grateful for useful comments on the manuscript to Joseph Barker as well as insightful discussions with Scott Bender, So Takei, Gen Tatara, Oleg Tretiakov, and Jiadong Zang. This work was supported by the US DOE-BES under Award No. DE-SC0012190 and in part by the ARO under Contract No. 911NF-14-1-0016 (S.K.K. and Y.T.) and by the US DOE-BES under Award No. DE-FG02-08ER46544 (O.T.).
\end{acknowledgments}

\bibliographystyle{C:/Users/evol/Dropbox/School/Research/apsrev4-1-nourl}
\bibliography{C:/Users/evol/Dropbox/School/Research/FDT-AFM}

\end{document}